\documentclass[12pt]{article}

\usepackage{epsfig}
\usepackage{graphicx}
\usepackage{color}

\def \d {\rm d}
\def \F {F}
\def \FF {\tilde F}
\def \A {\cal A}

\title{
On the Tail of the Overlap Probability Distribution in
the Sherrington--Kirkpatrick Model}

\author{
Alain Billoire\\
\small{Service de Physique Th\'eorique, CEA/DSM/SPhT} \\
\small{Laboratoire associ\'e au CNRS} \\
\small{CEA-Saclay, 91191 Gif-sur-Yvette c\'edex (France)}
\and
Silvio Franz\\ 
\small{ICTP, Strada Costiera 11, PO Box 563}\\
\small{34100 Trieste (Italy)}
\and 
Enzo Marinari\\
\small{Dipartimento di Fisica, SMC and UdR1 of INFM and INFN} \\
\small{Universit\`a di Roma {\em La Sapienza}}\\
\small{P. A. Moro 2, 00185 Roma (Italy) }
\and
\footnotesize{\tt billoir@spht.saclay.cea.fr}
\and
\footnotesize{\tt franz@ictp.trieste.it enzo.marinari@roma1.infn.it}
}

\begin{document}

\maketitle
\abstract{ We investigate the large deviation behavior of the overlap
probability density in the Sherrington--Kirkpatrick model using the
coupled replica scheme, and compare with the results of a large scale
numerical simulation. In the spin glass phase we show that generically,
for any model with continuous RSB, $1/N \log P_N(q)$ $\approx$ $- {\cal
A}$ $\left((|q|-q_{EA}\right)^3$, and compute the first correction to
the expansion of $\A$ in powers of $T_c-T$ for the SK model.  We also
study the paramagnetic phase, where results are obtained in the
replica symmetric scheme that do not involve an expansion in powers of
$q-q_{EA}$ or $T_c-T$. We give finally precise semi-analytical estimates of
$P(|q|=1)$.  The overall agreement between the various points of view
is very satisfactory. }

\section{Introduction\label{S-INT}}

Although the solution of the Sherrington--Kirkpatrick (SK) model was
proposed more than two decades ago \cite{Pa79}, and it physical
interpretation is (now) quite clear \cite{MPV87}, some fundamental aspects of the model are
still poorly understood.   We are
thinking here about the problem of the time scales of the model
\cite{dyn_paper}, and to the question of chaos with respect to
temperature variations \cite{chaos_1, chaos_2}, just to quote some
recent works.

In this article, we will revisit the question of the large deviation
behavior of the order parameter probability distribution $P_N(q)$ of the SK model
\cite{fpv,Pa93a, Pa93b}, using new analytical results, much
improved numerical data and refinements in the data analysis.

In the usual mean field approximation the probability distribution of
the order parameter $P_N(m)$ is read from the free energy considered
as a function of the order parameter. This method cannot be pursued in
the {\em replica approach}, since the replica free energy has no
physical interpretation for values of the order parameter different
from the saddle point \cite{MPV87}.  The {\em coupled replica method}
was proposed in \cite{fpv} to circumvent this problem and used in
\cite{Pa93a, Pa93b}.  In short, the replica method is applied to a
system of two identical copies of the original system with a
constrained value of the mutual overlap $q$, with the result:

$$ \frac1{N} \log
P_N(|q|>q_{EA}) = -{\A} \left(|q|-q_{EA}\right)^3 +{\cal O}
\left(\left(|q|-q_{EA}\right)^4\right)\ , $$ 

where $q_{EA}$ is the Edwards--Anderson order parameter (the maximum
allowed value for the overlap in the infinite volume limit). The
proportionality coefficient $\A$ was computed in \cite{fpv} to zeroth
order in $\tau=({T_c^2-T^2})/{2T_c}$.  In order to obtain a value of
the coefficient ${\A}$ to compare with the results of our large scale
simulation using the parallel tempering algorithm, we compute here the
first order
term in $\tau$.
We also present semi-analytical results for $P(|q|=1)$.  For $T\ge
T_c$ we present the results of an exact calculation, made in the
replica symmetric scheme, that is limited neither to small $q$'s nor
to the vicinity of $T_c$. This calculation corroborates nicely the
other analytical and numerical results.

There are somehow two aspects in the comparison made between the analytical
results and the Monte Carlo data. The first is to (cross) check both
results.  This is important since on one hand the analytical
computation is very delicate and involves some unproven
assumptions. The simulation is, on the other hand, limited to small
systems  finite number of disorder samples and finite statistics.  Both
approach are very complementary. The second aspect is in checking how much
of the infinite volume physics (at low temperatures, in the spin glass
phase) is already encoded in the sizes we can effectively study on our
current state of the art computers (up to $N=4096$ in the present
case): we will see that, as far as the present computation is
concerned, things do work well, and we are moving on solid ground.

\section{Large Deviations in $P_N(q)$ for $|q|>q_{EA}$\label{S-ANA}}

In this section\footnote{In the following we will take, without loss
of generality, $q>0$ (assuming  for example  the presence of a suitable 
infinitesimal magnetic field).} we sketch the main steps and give the
results of our computations of the function $P_N(q)$ in the large
deviation regime, i.e. for $|q|>q_{EA}$, for both $T<T_c$ and $T\geq T_c$.

The overlap probability distribution  $P_{N,J}(q)$
depends on the  disorder sample$J$, and the function  $P_N(q)$ is obtained as
$\overline {P_{N,J}(q)}$, where
the over-bar  denotes the disorder average. 
It is well known that in the thermodynamic limit the disorder average
of $P_{N,J}(q)$ is non-vanishing\footnote{We assume that our starting
Hamiltonian is symmetric under global spin reversal.} in the interval
$[0,q_{EA}]$, while sample to sample fluctuations remain strong even
for large systems \cite{MPV87}.  The
probability of the complementary interval, i.e. of events with
$q>q_{EA}$, is exponentially small in the system size \cite{fpv}. For
typical samples one has, independently of the disorder
realization,
\begin{equation}
P_{N,J}(q)\approx 
\tilde{
P}_N(q) \propto \exp \left( -\beta N  F(q) \right)\ ,
\end{equation}
where $\beta=1/{T}$ and $ F(q)$ is the self-averaging
free energy cost of keeping two replicas at a mutual overlap $q$:
\begin{equation}
   F(q)=-\frac{T}{N}\overline {\log\left( \frac{1}{Z^2}
  \sum_{\{ \sigma_i,\tau_i\}}
  e^{-\beta \left(H(\sigma)+ H(\tau)\right)}
  \ (q-\frac{1}{N}
\sum_i \sigma_i\tau_i)\right)}\ .
  \label{F2}
\end{equation}
It is quite natural to expect that in the thermodynamic limit
$P(q)=\tilde{P}(q)$. In an alternative scenario the fluctuations of
$P_{N,J}(q)$ are dominated by rare (i.e. having exponentially
vanishing probability) samples, causing $\overline{P_{N,J}(q)}$ to be
different from $\tilde{P}_N(q)$. We will discuss this issue with our
numerical simulations in the next section.

Let us start with the case $T<T_c$.  In reference \cite{fpv} the
computation of $ F(q)$ has been performed using the method of {\em
coupled replicas} in the glassy phase ($T<T_c$) close to $T_c$, with
the result that, in leading order in both $\tau\equiv
({T_c^2-T^2})/{2T_c}$ and $q-q_{EA}$, on has $ F=1/{6}(q-q_{EA})^3$.
Let us now perform the computation of the prefactor of the cubic term
to the next order in $\tau$.
In order to compute $ F(q)$ one has to replicate $n$ times both
the spins $\sigma_i$ and the spins $\tau_i$. Two order parameter
matrices appear:
\begin{equation}
Q_{ab}=\langle \sigma_a \sigma_b\rangle=\langle \tau_a \tau_b\rangle
\;\;\mbox{and}\;\;
P_{ab}=\langle \sigma_a \tau_b\rangle\;\;\ ,\;\; a,b=1,...,n\ .
\end{equation}
One makes a Parisi ansatz \cite{MPV87} for both $Q$ and $P$,
introducing two functions
$q(x)$ and $p(x)$.  
We now recast the two $n\times n$ matrices into a single
$2n\times 2n$ bold-faced matrix
\begin{equation}
{\bf Q}=\left(
\begin{array}{ll}
Q & P\\
P & Q
\end{array}\right)\ .
\end{equation}
Let us consider in all generality any system that is described in the
un-coupled case by the free energy functional $F[Q]$ in terms of the
usual replica matrix. The same system when coupled according to 
(\ref{F2}) has a free energy functional of the form
\begin{equation}
F_2[Q,P]=F[{\bf Q}]+\epsilon(nq-\sum_{a=1}^n P_{aa})\ ,
\end{equation}
where $\epsilon$ is a Lagrange multiplier associated to the
$\delta$-function in (\ref{F2}).

If the problem of a single, un-coupled, system admits a continuous
solution $q_F(x)$ (that we suppose to be known), it can be shown that the
corresponding variational equations related to the coupled problem
admit two simple solutions \cite{tesi}.

A first solution has $\Delta F=\epsilon=0$, with  functions $q(x)$
and $p(x)$ that can be constructed explicitly from $q_F(x)$:
\begin{equation}
q(x)=\left\{
\begin{array}{ll}
q_F(2 x)& 0 \leq x  \leq \frac{\tilde{x}}{2}\\
q &  \frac{\tilde{x}}{2} \leq x < \tilde{x}\\
q_F(x)& 1 \ge  x \ge \tilde{x}
\end{array}
\right. \;\;\;\;
p(x)=\left\{
\begin{array}{ll}
q_F(x)& 0  \leq  x  \leq {\tilde{x}}\\
q &  {\tilde{x}} \leq x< 1\ ,
\end{array}\right.
\end{equation}
where the point $\tilde{x}$ is defined by the equation $q_F(\tilde{x})
= q$.

In the second solution $q(x)=p(x)$. This second solution becomes
degenerate with the first one at $q=q_{EA}$.  If we take the usual
replica approach attitude to {\em maximize} the free energy with
respect to the $Q$-parameters (in agreement with the physical
intuition about the problem) we will select the first solution in the
region with $q< q_{EA}$ and the second one in the region with $q>
q_{EA}$, where it has a larger free energy.

Notice that while the first solution has a flat free energy $
F(q)=0$, the second, as a consequence of the facts that 

\begin{itemize}
\item 
it coincides with the first one for $q=q_{EA}$; 

\item 
it has to be a solution of the saddle point equations; 
\end{itemize}

\noindent
must be such that $\frac{\d  F(q)}{\d q}|_{q=q_{EA}}=0$.  All
these properties together imply that generically $  F(q)={\A}
(q-q_{EA})^3$, where ${\A}$ is a problem and temperature
dependent constant. Of course the accidental vanishing of $\A$ is an
allowed possibility, and in that case $  F(q)$ would be of order
five or higher.

Let us now turn to the explicit computation for the SK model close to
$T_c$. In this case the replica free energy density can be written as
\begin{equation}
F_2[Q,P]=\tau\ Tr {\bf Q^*}^2+\frac{1}{3}Tr {\bf Q^*}^3 +\frac{1}{4}y
\sum_{\alpha,\beta}^{1,2n} {\bf Q^*_{\alpha,\beta}}^4
+\epsilon(nq-\sum_{a=1}^n P^*_{aa})\ ,
\end{equation}
where $y=2/3$, but will be  kept  as a parameter 
during the computation, and we have defined
$Q^*_{ab}\equiv {Q_{ab}}/{T^2}\simeq Q_{ab}(1+2\tau)$ and
$P^*_{ab}\equiv {P_{ab}}/{T^2}$. Here the solution can be found
explicitly,   writing $\tilde{p^*}=P^*_{aa}$ and inserting
$P^*_{ab}=Q^*_{ab}$ ($a\ne b$) in the equation of motion one finds that
the matrix $Q^*$ verifies the equation:
\begin{equation}
(\tau+\tilde{p}^*)Q^*_{ab}+(Q^{*2})_{ab}+\frac{y}{2}\ Q_{ab}^{*3}=0\ ,
\end{equation}
which is identical to the solution of the free problem with $\tau\to
({\tau+\tilde{p^*}})/{2}$ and $y\to {y}/{2}$. Plugging then the
known solution of the free problem into the free energy functional it is easy to derive
the result:
\begin{equation}
 F(q)=\left(\frac {1}{6} -\frac {3}{4} \tau y\right)(q^*-q^*_{EA})^3.
\label{Ana}
\end{equation}
Or, using the original definition of $q$, and setting $y=2/3$,
\begin{equation}
\label{Ana_two}
 F(q)= \left(\frac16+\frac12\tau\right)\left(q-q_{EA}\right)^3\ .
\end{equation}

Let us now discuss the paramagnetic phase, where $T>T_c$.  Here
$q_{EA}=0$ and we have to select the  solution that has always
$p(x)=q(x)$ for $q\ge 0$ and $p(x)=-q(x)$ for $q< 0$.  Let us consider
the replica symmetric free energy, corresponding to $\tilde{p}=q$,
$q(x)={\rm sgn} (q), p(x)=q_0$. For simplicity we will write as before
equations valid for $q>0$, keeping in mind that the free energy
$F(q)$ has to satisfy $F(-q)=F(q)$.

Defining $Dy= {{\rm e}^{-\frac{1}{2}y^2}}/{\sqrt{2\pi}} dy$,
standard manipulations lead to the expression
\begin{eqnarray}
\nonumber
  F(q) &=& -\frac{\beta}{2} (1+q^2+2q_0^2-2q_0-2 q q_0 )
    -T \log 2+T\nu q\\
    &-& T\int Dy
    \log\left[ {\rm e}^{-\nu}
    +{\rm e}^{\nu}\cosh (2\beta\sqrt{q_0} y )\right] \ ,
\label{RS_1}
\end{eqnarray}
that has to be extremized with respect to $q_0$ and
$\nu$ (the Lagrange multiplier associated to $q$). This 
leads to the saddle point equations
\begin{eqnarray}
q&=&\int Dy \frac{- {\rm e}^{-\nu}+{\rm e}^{\nu}\cosh (2\beta\sqrt{q_0} y)
}{{\rm e}^{-\nu}+{\rm e}^{\nu}\cosh (2\beta\sqrt{q_0} y)},\\ q_0&=&
\frac{1+q}{2}-\frac{1}{2\beta \sqrt{q_0} y} \int Dy 
\frac{y {\rm e}^{\nu}\sinh (2\beta\sqrt{q_0} y) }{{\rm e}^{-\nu}+{\rm
e}^{\nu}\cosh (2\beta\sqrt{q_0} y ) }.
\end{eqnarray}
These equations have always the trivial solution $q_0=0$, $\nu=\tanh^{-1}q$,
which is correct at high enough temperatures and small values of
$q$. A small $q_0$ expansion reveals that a solution with $q_0\ne 0$
is possible for $q\ge T-1=q_c(T)$. Notice that
\begin{itemize}
\item
for $T\le 1$ the solution is non-trivial for all  values of $q$,
\item
for $T> 2$ the solution is always $q_0=0$, 
\item
for $1<T\le 2$ there is a phase transition from the solution $q_0=0$ for $q<T-1$ to 
a solution $q_0\ne 0$ for $q>T-1$.
\end{itemize}
The correct solution in this last domain should break the replica
symmetry. Close to $q_c(T)$ a small $q$ expansion as in the previous
section can be used even for large values of $T_c-T$. For
$q_c(T)-q\approx 1$ we only consider the replica symmetric
approximation. It is clear that if we impose $q=1$ the system behaves
as a single system with temperature ${T}/{2}$, and we understand
therefore why $q_c(2)=1$.

\begin{center}
\begin{figure}
\epsfxsize=250pt \epsffile{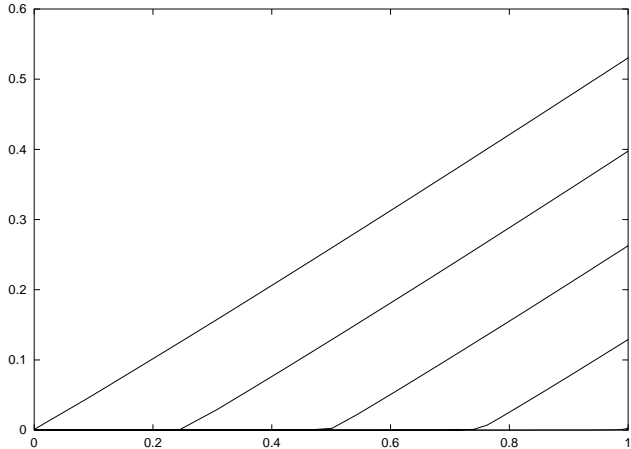}
\caption[0]{The induced overlap $q_0$ as a function of the mutual
overlap $q$ that we force on the system for $T=1.0,1.25,1.5,1.75$
and $2.00$
(from right to left, with  $q_0=0$ $\forall q$ for $T=1$). 
\label{fig1a}   
}
\end{figure} 
\end{center}

\begin{center}
\begin{figure}
\epsfxsize=250pt \epsffile{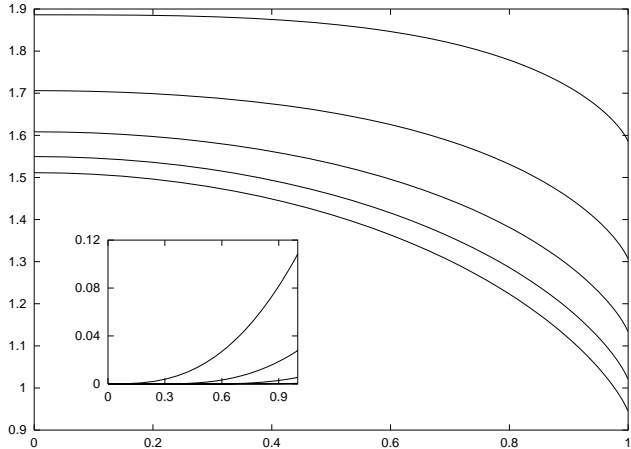}
\caption[0]{The free energy $-\beta F(q)$ for the same values of the
temperature as in figure \ref{fig1a} (from top to bottom). In the
insert we plot, for the same set of temperatures again,
the difference $-\beta (F_0(q)-F(q))$ , where $F_0(q)$ is
computed with the trivial solution $q_0=0$ (from top to bottom. The
difference is zero for $T=2$ and very small for $T=1.75$).
\label{fig1b}   }
\end{figure} 
\end{center}

The $q_0=0$ solution gives a free energy
\begin{equation}
F(q)=-\frac{\beta}{2} (1+q^2)-2 T \log 2+T\nu q-T
\log(\cosh (\nu))\ .
\end{equation}
In figure (\ref{fig1a}) we show the solution for $q_0$ as a function
of $q$. For $q=1$ we find that $q_0$ takes, as it should, the value 
$q_{RS}(T/2)$, i.e.  the Edwards--Anderson value at temperature
${T}/{2}$ in the RS approximation.  The functional form of $q_0$
is well approximated for all values of $T$ by $q_0(q)=
q_{RS}({T}/{2}) (q-q_c(T))$, although small quadratic deviation
from this form are observable. In figure (\ref{fig1b}) we also plot
the related free energy.

\section{Numerical versus Analytic Results\label{S-NUM}}

In the following we will use our numerical results obtained from
simulations of the $J_{i,j}=\pm 1$ SK model toward two different goals, that we have
already partially discussed in the introduction. First we want to have
a numerical check of our analytical findings: it is clear that we are
dealing with complex set of equations, we are working in some
asymptotic regimes, and  this implies that a numerical check is quite
welcome. The second issue is maybe less direct but also very
crucial. It is important to understand how well numerical simulations
done on finite lattice encompass the infinite volume physics, and, at
the same time, how good a control we can have on the temperature range
that we explore. The issue of the infinite volume limit is indeed of
very large importance, and it is crucial to verify that exotic
phenomena (like the large deviation regime we are discussing here) are
already well quantified in the region of lengths we can study with
nowadays computer facilities.

Our large scale numerical simulations are based on the numerical
optimized Monte Carlo technique of  {\em parallel tempering}
\cite{MARINA}.  They  have already
been used in our recent paper on the problem of chaos \cite{chaos_2}
(but for the addition of the data obtained on a $N=128$ site lattice).

We have analyzed $1024$ different disorder realizations for $N=64$,
$256$ and $1024$ sites, $256$ for $N=4096$ and $8192$ for $N=128$. We
go down to a minimal temperature value $T_m=0.4\ T_c=0.4$, and we use
$0.4\ 10^6$ iterations for thermalizing and $10^6$ iterations for
measurements.  We normalize probability distributions according to
$\int_{-1}^1 P_N(q) dq = 1$.

\begin{figure}
  \centering 
  \includegraphics[width=0.5500\textwidth,angle=270]
{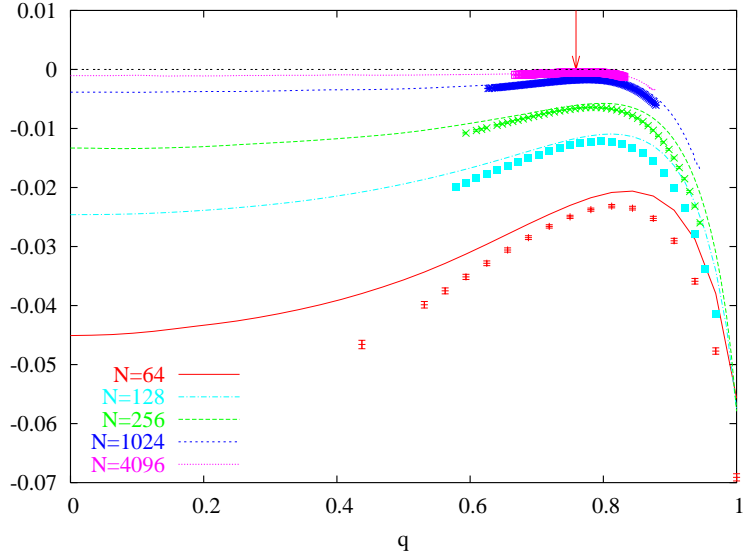}
\caption{ ${1}/{N} \log\left(P_N(q)\ N^{-\frac13}\right)$ versus
$q$ for $T=0.4$ : annealed ($\F(q)$, represented by lines) and
quenched ($\FF(q)$, represented by points with error bars)
estimates. The vertical arrow indicates the value of $q_{EA}$ at
$T=0.4$.}
\label{one}
\end{figure}

The first question we discuss is whether it is appropriate to measure
the quench\-ed average 
\begin{equation}
  \beta{  \FF(q)} = N^{-1}\ {\overline{
  \log P_{N,J}(q)}},
  \label{E-QUENCHED}
\end{equation} 
(that is the quantity computed using the two-replica method)
or if we can be happy by measuring the annealed average 
\begin{equation}
  \beta{
  {\F}(q)} = N^{-1}\ \log {\overline{ P_{N,J}(q)}},
  \label{E-ANNEALED}
\end{equation} 
(a quantity that
is much easier to measure with Monte Carlo simulation).  In section
\ref{S-ANA} we have discussed the issue, arguing that these two
averages could differ because of rare samples. We now turn to our numerical
data in order to answer this question.

In figure \ref{one} we show results for both the quenched (equation
\ref{E-QUENCHED}) and the annealed (equation \ref{E-ANNEALED})
averages at $T=0.4$ as a function of $q$. We plot data for $N=64$,
$128$, $256$ and $1024$.  In order to make the figure readable:

\begin{enumerate}
\item 
we have subtracted from the different measurements a term $1/3 \log
N$, that  separates the points from different system sizes;

\item 
we have not drawn the $F(q)$ data as points with statistical error
bars,  but we have drawn a line through the data points.
\end{enumerate}

\noindent
The (few) data points for $\FF(q)$ are drawn with statistical error
bars (estimated from the variance of sample to sample fluctuations).
It appears that the data for $\FF(q)$ are confined to a much smaller
$q$ range than the data for $\F(q)$. This is expected since for a
given value of $q$, $ {\FF(q)}$ is well defined only if (the estimate
of) $P_N(q)$ is larger than zero for every sample, whereas $\F(q)$ is
well defined as soon as the average $P_N(q)$ is larger than zero.

Figure \ref{one} shows that for large system sizes the annealed and
quench\-ed estimates are consistent. The same conclusion holds for all
the other values of the temperature that we have considered. Because
of that we will only consider in what follows the annealed averages
$P_N(q)$ and $F(q)$. We will estimate statistical errors on our
observables by a jack-knife procedure \cite{FLYVBJ}.

\begin{figure}
  \centering 
  \includegraphics[width=0.5500\textwidth,angle=270]{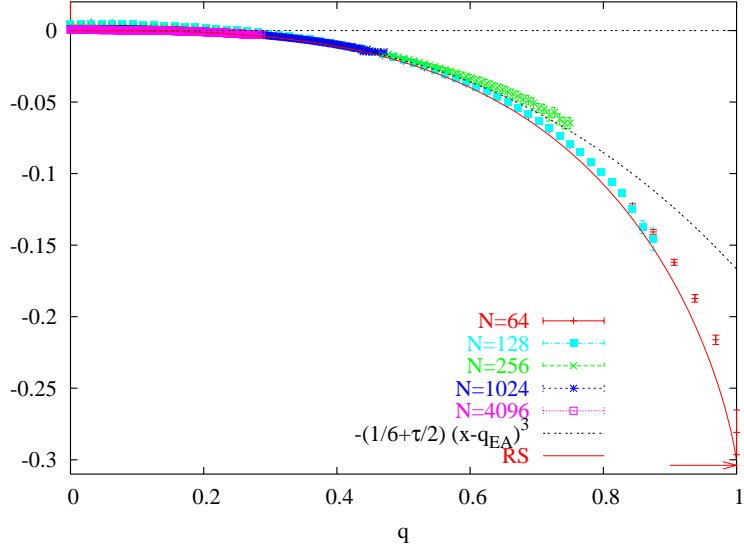}
\caption{${1}/{N} \log P_N(q)$ versus $q$ for $T=1$: points with
error bars are numerical data from different lattice sizes, the dotted
line is from the perturbative estimate in the spin glass phase.  An
horizontal arrow indicates the exact $q=1$ value. The full line
(labeled ``RS'') is from equation \ref{RS_1}.}
\label{two}
\end{figure}

Our second (and main) numerical goal is to verify the predictions of
equation \ref{Ana_two} that predict asymptotically a cubic behavior of
$1/N \log P_N(q)$ in $q-q_{EA}$, with a coefficient computed at order
$\tau$. Now, is all that correct? Can we already observe this behavior
on the lattice size that we are able to thermalize in the spin glass
phase?

\begin{figure}
  \centering 
  \includegraphics[width=0.5500\textwidth,angle=270]{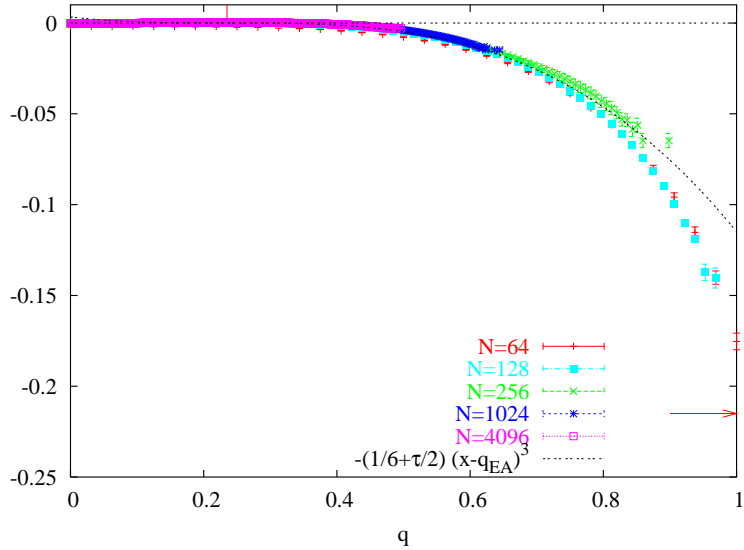}
\caption{As in figure \ref{two} but with $T=0.8$.  
The additional vertical arrow indicates the value of $q_{EA}$.}
\label{three}
\end{figure}

\begin{figure}
  \centering 
  \includegraphics[width=0.5500\textwidth,angle=270]{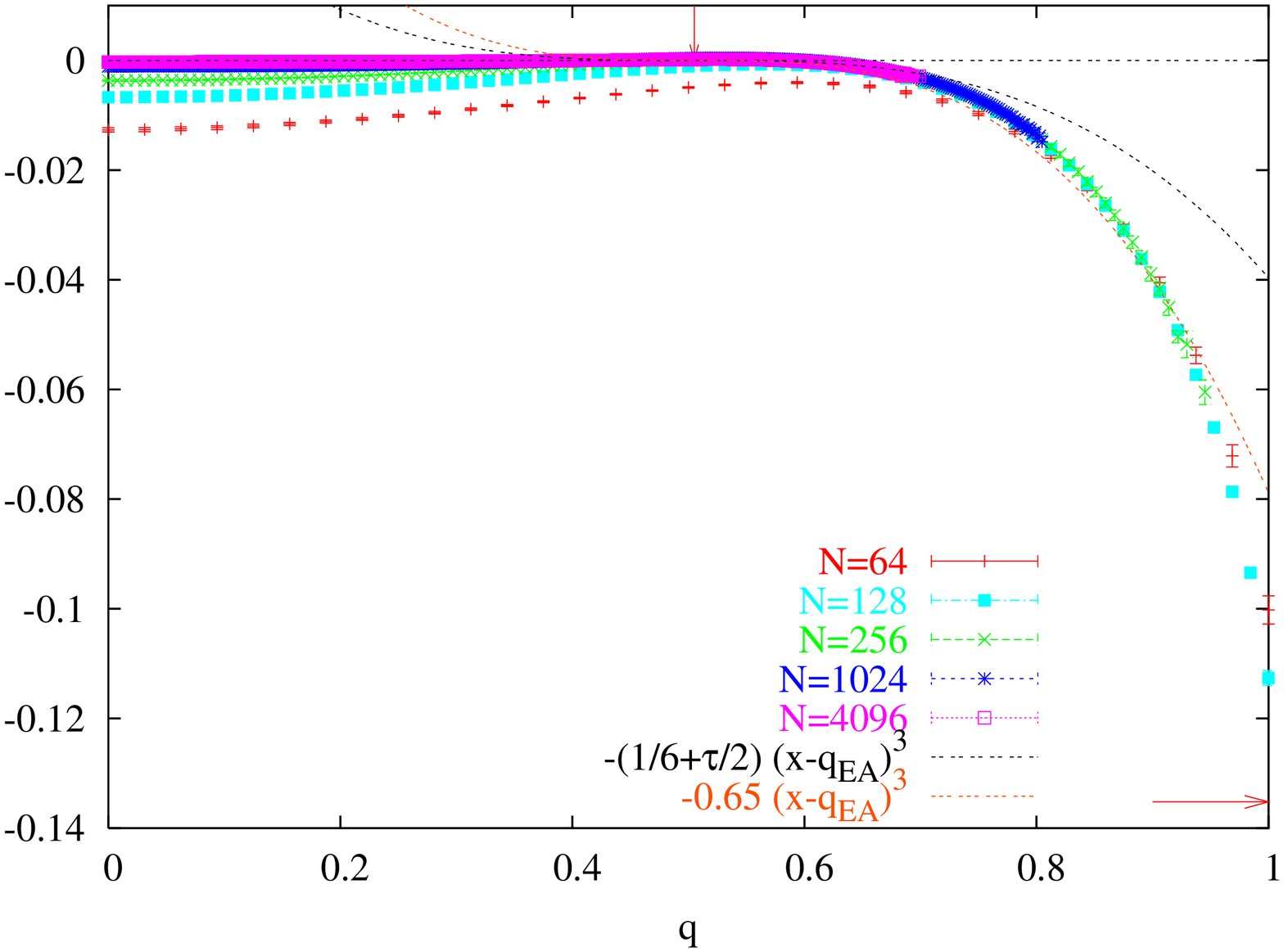}
\caption{As in figure \ref{three} but with $T=0.6$.  The additional lower
dotted curve is drawn using a coefficient modified by hand,
accounting for the renormalization of the temperature when $\tau$
grows.}
\label{four}
\end{figure}

\begin{figure}
  \centering 
  \includegraphics[width=0.5500\textwidth,angle=270]{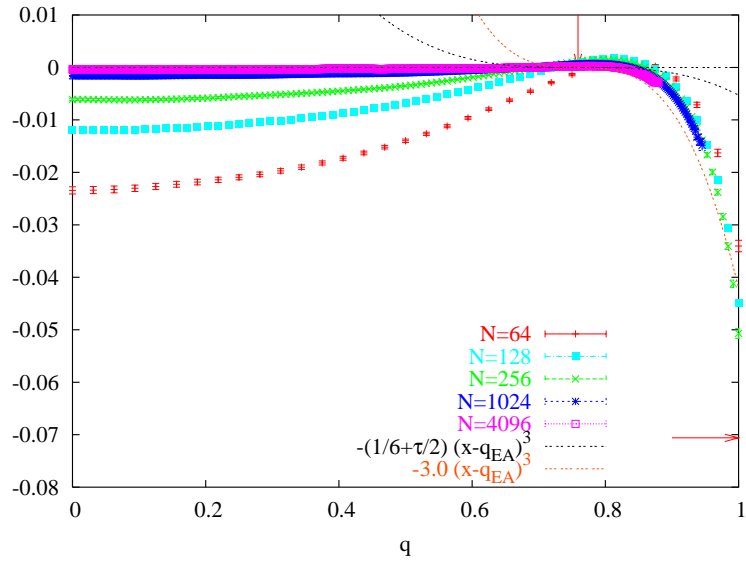}
\caption{As in figure \ref{four} but with $T=0.4$.}
\label{five}
\end{figure}

We report our results at different values of the temperature ($T=1.0$,
$0.8$, $0.6$ and $0.4$) in figures \ref{two},
\ref{three}, \ref{four} and \ref{five}  respectively.  With the dotted line 
we plot the theoretical prediction of equation \ref{Ana_two}. With an
horizontal arrow we give the analytic value of $P_N(q=1)$ (see the
discussion later on).

\begin{figure}
  \centering 
  \includegraphics[width=0.5500\textwidth,angle=270]{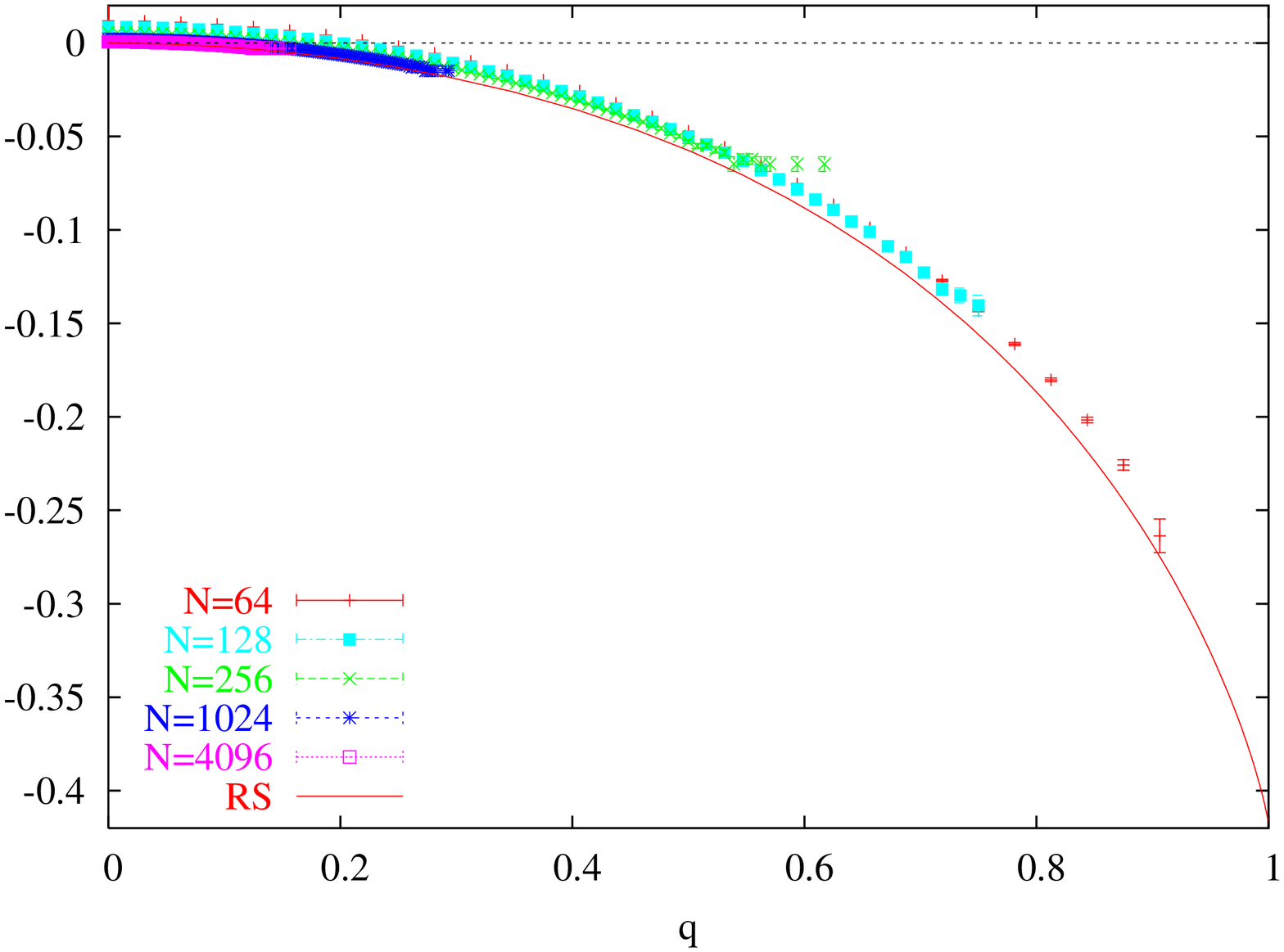}
\caption{As in figure \ref{four} but with $T=1.3$.}
\label{zero}
\end{figure}

All figures show that results for the systems of different sizes are
converging to a common ($N$ independent) function: our results for the
excess free energy are indeed in the asymptotic regime.  The careful
reader will notice however that only the limit $\lim_{\to \infty} 1/N
\log P_N(q)$ is predicted. There is accordingly an order $1/N$
ambiguity in the vertical offsets of our plots for $1/N \log P_N(q)$.
Plotting for example $1/N \log \left( P_N(q)N^{-1/3}\right)$ would
result in a slightly different looking figures.

At $T=1$, where $q_{EA}=0$, the agreement of our data with the
theoretical prediction is very good at least up to $q=0.6$.  Also at
$T=0.8$, where $q_{EA}$ is slightly larger than $0.2$ (i.e.  still
very small) the agreement is very good in a very large $q$ range (up
to $q\simeq 0.8$).  In all these analysis we have used the value of
$q_{EA}$ computed by Crisanti and Rizzo by numerical integration
\cite{Crizzo} of the Parisi solution of the SK model (e.g. we are not using 
$q_{EA}$ computed to order $\tau$).

At $T=0.6$, where $q_{EA}$ is of order $0.5$ and a small $\tau$
expansion with two terms starts to be inappropriate, the agreement
starts to be less good. To take into account the fact that here $\tau$
is large and there is an effective renormalization (when $T=0.6$ the
first order correction in equation \ref{Ana_two} is equal to $0.16$
while the leading term is equal to $1/6$) we have added a further
curve where we substitute the prefactor we have computed analytically
in equation \ref{Ana_two} with some effective value, that optimizes
the matching of the analytic form to the numerical data: this improves
very much the agreement.  This effect is even more dramatic at
$T=0.4$, where again a cubic dependence with a renormalized prefactor
fits very well the data in a large $q>q_{EA}$ range.

Our third numerical goal concerns the value of
of $P_N(q)$ at $q=1$ (that we plot with an horizontal arrow in the
four figures). We use the simple relation
\begin{equation}
  \frac{1}{N} \log P_N(q=1) = - \frac{2}{T} 
  ( {\cal F}_{\frac{T}{2}}-{\cal F}_{T}) \ ,
\label{TRUE-ONE}
\end{equation}
between $P_N(q=1)$ at temperature $T$ and the free energy 
of the SK model at temperature $T$ and $T/2$. We take for
${\cal F}_T$ the infinite volume free energy of the model
derived with high precision in \cite{Crizzo} from Parisi solution.

One sees from figures \ref{two}-\ref{five} that the prediction of equation
\ref{TRUE-ONE} for $P_N(q=1)$ is smaller than the value that formula
\ref{Ana_two} takes at $q=1$: obviously this had to be expected, since
equation \ref{Ana_two} is the first term in an expansion in powers of
$q-q_{EA}$, and is not supposed to be valid up to the $q=1$ boundary,
where higher orders cannot be neglected (the function is presumably
singular  at $q=1$) .
Note however that the interplay between finite size effects and higher
orders in $T_c-T$ is not trivial. It could be studied numerically
using the multi-overlap algorithm \cite{BBJ} which allows to measure
$P_N(q)$ with high accuracy up to $q=1$. However parallel tempering
numerical data, when pushed to very high statistics, confirm already
in a remarkable way the analytic value (this can be only done for
small lattice sizes). At $T=1$ we are able to push the $N=64$ lattice
up to $q=1$ and to keep under control the limit of the $N=128$
lattice. At lower $T$ values the check is slightly more difficult, but
one sees that the $q=1$ limit of larger and larger lattices approaches
better and better the analytic value.

\begin{figure}
  \centering 
  \includegraphics[width=0.5500\textwidth,angle=270]{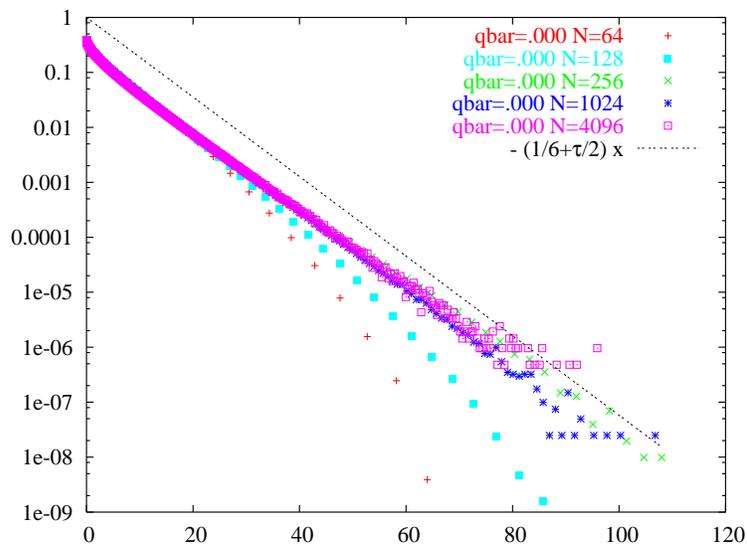}
\caption{Scaling plot of $P_N(q)\ N^{-\frac13}$ as a function of
$(q-q_{max}(N))^3 N$ for $T=1.0$, compared with the coupled replica
estimate (with arbitrary normalization).  Our estimate of $q_{max}(N)$
is reported in the figure under the name {\bf qbar} (it is obviously
zero at $T_c$) .}
\label{six}
\end{figure}

\begin{figure}
  \centering 
  \includegraphics[width=0.5500\textwidth,angle=270]
{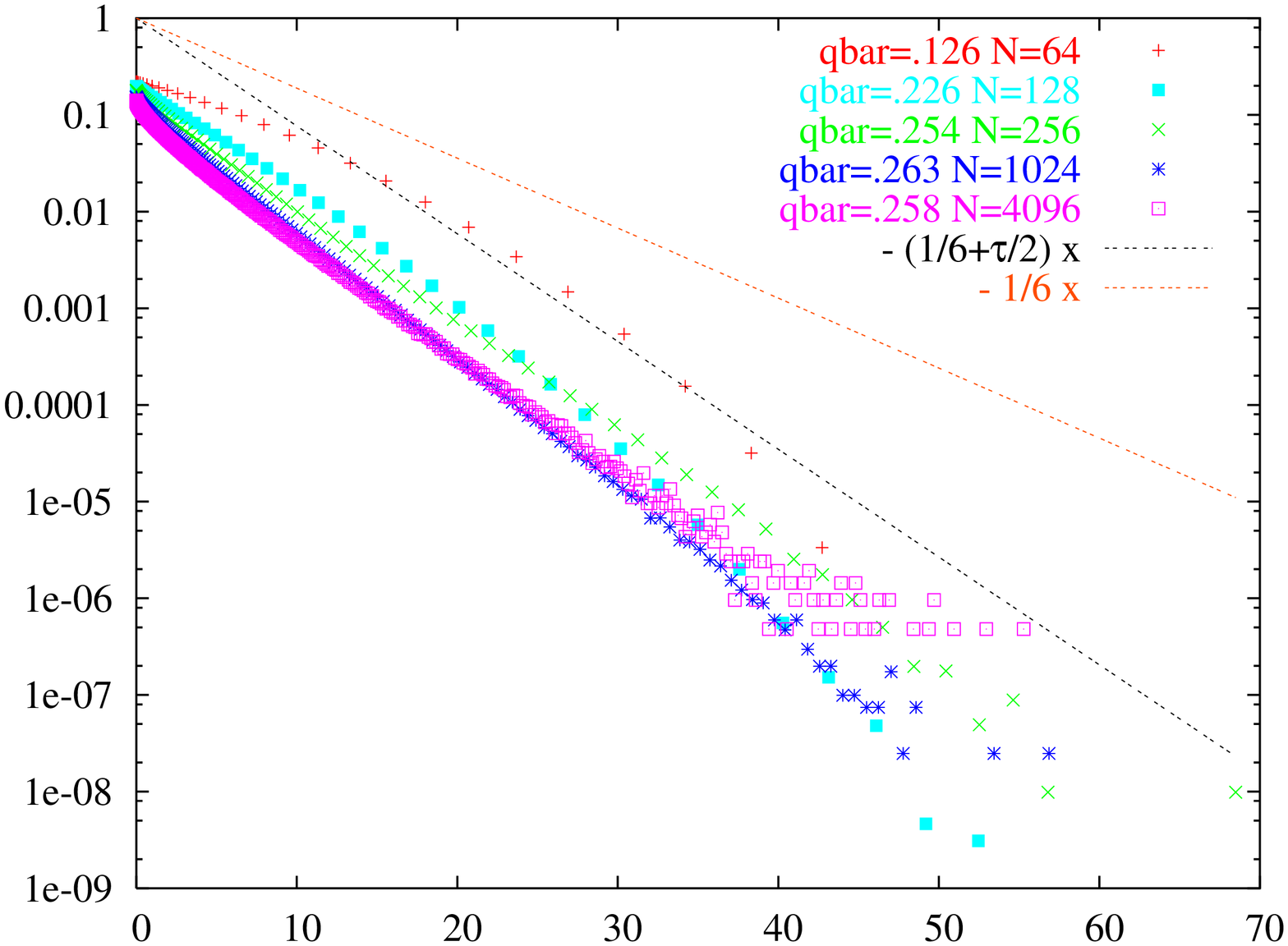}
\caption{As in figure \ref{six} but for $T=0.8$.  
Coupled replica estimate is plotted with and without the term linear in
$\tau$.  Note that $q_{max}(N)$ is not well determined on small
systems at this temperature, that is close to $T_c$.}
\label{seven}
\end{figure}

That the situation is well under control can be shown nicely for $T
\geq 1$ by comparing the numerical data, in the whole $q\in [0,1]$
range, with the results of our RS computation of equation
\ref{RS_1}. This is done in figures \ref{two} for $T=1.0$ and
\ref{zero} for $T=1.3$ (the highest temperature alas in our parallel
tempering simulation). The first figure shows the full line (labeled
``RS'') cleanly overshooting the dotted line, to reach the $q=1$ axis
slightly above (as expected) the exact $q=1$ limit\footnote{This
strongly indicates that the correct large $q$ behavior in figures \ref{two}
and \ref{three} is given by the $N=64$ and $128$ data, and that the
rightmost $N=256$ points are somewhat misleading}. Figure
\ref{zero} show also again the agreement our our numerics with the
replica calculation.

Equation \ref{Ana_two} suggests \cite{Pa93a,Pa93b} that (for $T<T_c$)
$P_N(q)$ scales asymptotically as
\begin{equation}
  P_N(q)\ N^{-\frac13} = 
  {\cal F}\left(N\left(q-q_{EA}\right)^3\right) 
  \qquad \qquad\mbox{for}\ \  q>q_{EA}\ ,
\label{scaling_F}
\end{equation}
with ${\cal F}(z) \propto \exp(-A z)$, with some positive constant
$A$, for large $z$.  If the above formula holds down to the maximum of
$P_N(q)$ this implies that the location of the peak of the probability
distribution of the overlap on a lattice of size $N$, that we denote
by $q_{max}(N)$, has to behave like
$$
  q_{max}(N) = q_{EA}+ c N^{-\frac13}\ ,
$$
a behavior that has indeed been verified with high accuracy in former work
\cite{chaos_2}.

\begin{figure}
  \centering 
  \includegraphics[width=0.5500\textwidth,angle=270]
{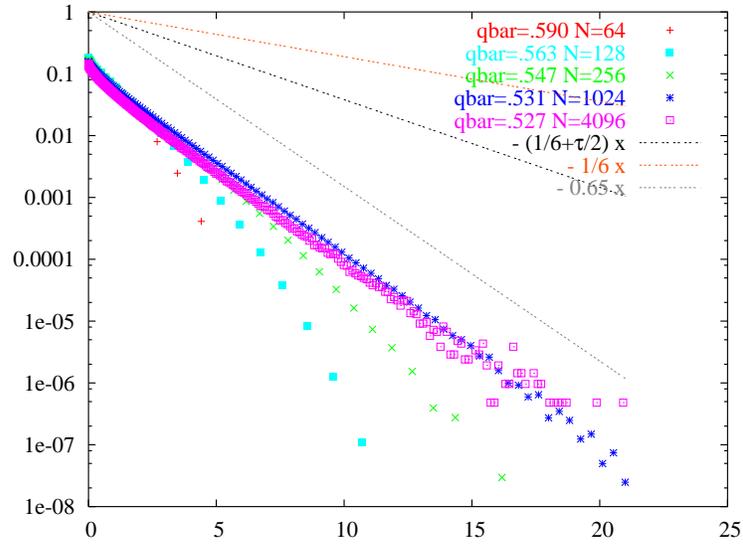}
\caption{As in figure \ref{seven} but for $T=0.6$.
The additional dotted line is drawn using the same hand modified
coefficient as in figure \ref{four}.}
\label{eight}
\end{figure}

\begin{figure}
  \centering 
  \includegraphics[width=0.5500\textwidth,angle=270]
{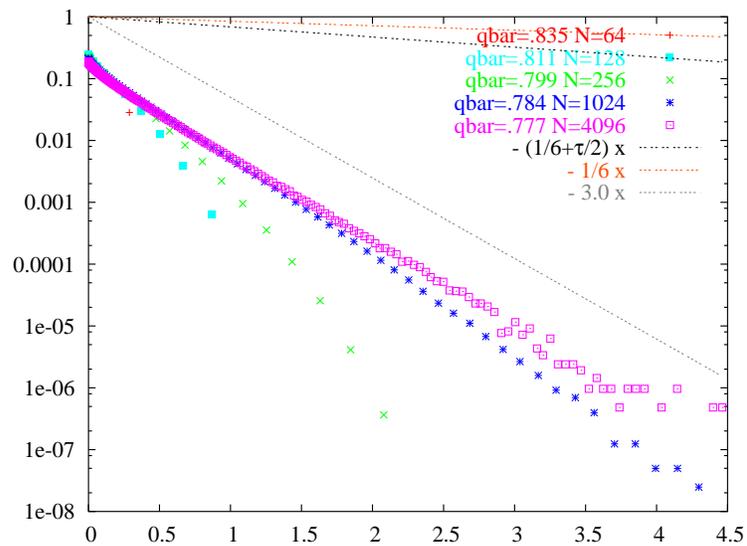}
\caption{As in figure \ref{six} but for $T=0.4$. The additional dotted line is drawn using the same hand modified
coefficient as in figure \ref{five}.}
\label{nine}
\end{figure}

In what follows we will accordingly look for a scaling law in the form
\begin{equation}
  P_N(q)\ N^{-\frac13} = 
  {\cal G}\left(N\left(q-q_{max(N)}\right)^3\right)
\label{scaling_G}
\end{equation}
using the numerical estimates obtained for $q_{max(N)}$.  Figures
\ref{six}, \ref{seven}, \ref{eight} and \ref{nine} are scaling plots
of $\log \left( P_N(q) N^{-1/3}\right)$ versus $N
(q-q_{max}(N))^3$. The figures clearly exhibit the predicted cubic
limiting behavior.  At $T=T_c$ the data are furthermore in excellent
agreement with the predicted slope.  At $T=0.8$ the term linear in
$\tau$ computed in this paper is essential in order to obtain good
agreement with the data. $T=0.6$ and $T=0.4$ are clearly too far from
$T_c$ in order to obtain a good result for the slope with only two
terms in the expansion in powers of $\tau$.  Scaling is violated on
small systems for the rightmost points. It must be so since those are
points with $q$ close to one, and we have seen before that close to
one the behavior is not cubic.

It should be stressed that with our data scaling in
$(q-q_{max}(N))^3 N$ (equation \ref{scaling_F}) is much better than scaling 
in $(q-q_{EA})^3 N$ (equation \ref{scaling_G}, using
the very precise data for $q_{EA}$ by Crisanti and Rizzo).    
One reason for this better agreement may
be that for $q>q_{EA}$ the inflection point of $(q-q_{max(N)})^3 N$
mimics  the maximum of $1/N \log P_N(q)$.
Scaling plots as function of $(q-q_{EA})^3 N$ with an ad-hoc effective
$N$ independent $q_{EA}$ appear (in a small $z$ interval) in
\cite{Pa93a} for $T=0.8$, and in \cite{Pa93b} for $T=0.5$.

\section{Conclusions\label{S-CON}}

The results of these analysis are very positive. We have first argued
that the large deviation leading behavior of $P_N(q>q_{EA})$ is
generically $1/N \log P_N(q)$ $=$ $- {\cal A}$
$\left(q-q_{EA}\right)^3$, and we have computed the first corrections
to the expansion of $\A$ in powers of $T_c-T$ for the
Sherrington--Kirkpatrick model.

Data from numerical simulations confirm that this behavior can be
detected already on moderate lattice sizes.  First we have seen that
we are well allowed to take an annealed average of $
F$. Secondly we have analyzed the details of the cubic behavior, and
checked the validity of the perturbative estimate of the prefactor:
close to $T_c$ it works very well, while far away from $T_c$ a simple
renormalization improves the agreement in a substantial way.  Our very
accurate data have allowed a detailed analysis of the $P(q=1)$ value,
allowing again for a positive check of the analytic result. For $T \ge
T_c$ an exact computation in the replica symmetric scheme shows nicely
the crossover from the $q \approx q_{EA}$ behavior to the ultimate
limit $q=1$. At last we have shown that the scaling plots for $P_N(q)$
improve appreciably if one uses $(q-q_{max}(N))^3 N$ as the scaling
variable, showing clearly the predicted cubic behavior of $\log \left(
P_N(q)\ N^{-1/3}\right)$.

\newpage

\section*{Acknowledgments} 
We thank A. Crisanti and T. Rizzo for providing us with their
unpublished numerical evaluations of quantities in the SK model, and
Giorgio Parisi for conversations.  SF thanks the SPhT Saclay for a
visiting professor fellowship and warm hospitality in the months of
October-December 2001, during which part of this work was performed.

\end{document}